\documentclass[fleqn,usenatbib]{mnras}

\usepackage{newtxtext,newtxmath}

\usepackage[T1]{fontenc}

\DeclareRobustCommand{\VAN}[3]{#2}
\let\VANthebibliography\thebibliography
\def\thebibliography{\DeclareRobustCommand{\VAN}[3]{##3}\VANthebibliography}

\usepackage{graphicx}	
\usepackage{amsmath}	

\title[Pegasus V]{Pegasus V - a newly discovered ultra-faint dwarf galaxy on the outskirts of Andromeda}

\author[Collins, M. L. M. et al. ]{Michelle L. M. Collins$^{1}$\thanks{E-mail: m.collins@surrey.ac.uk}, Emily J. E. Charles$^{1}$, David Mart{\'\i}nez-Delgado$^{2}$,
 Matteo Monelli $^{3,4}$,\newauthor Noushin Karim$^{1}$,
Giuseppe Donatiello$^{5}$,
Erik J. Tollerud$^{6}$, Walter Boschin$^{3,4,7}$
\\
$^{1}$Physics Department, University of Surrey, Guildford, GU2 7XH\\
$^{2}$Instituto de Astrof\'isica de Andaluc\'ia, CSIC, Glorieta de la Astronom\'\i a, E-18080, Granada, Spain \\
$^{3}$Instituto de Astrof\'isica de Canarias (IAC), Calle V\'ia L\'actea s/n, E-38205 La Laguna, Tenerife; Spain \\
$^{4}$Facultad de F\'isica, Universidad de La Laguna, Avda. Astrof\'isico Fco. S\'anchez s/n, 38200La Laguna, Tenerife, Spain. \\
$^{5}$UAI -- Unione Astrofili Italiani /P.I. Sezione Nazionale di Ricerca Profondo Cielo, 72024 Oria, Italy \\
$^{6}$ Space Telescope Science Institute,  3700 San Martin Drive, Baltimore, MD 21218, USA \\
$^{7}$Fundaci\'on G. Galilei - INAF (Telescopio Nazionale Galileo), Rambla J. A. Fern\'andez P\'erez 7, E-38712 Bre\~na Baja (La Palma), Spain
}

\date{Accepted XXX. Received YYY; in original form ZZZ}

\pubyear{2022}

\begin{document}
\label{firstpage}
\pagerange{\pageref{firstpage}--\pageref{lastpage}}
\maketitle

\begin{abstract}
We report the discovery of an ultra-faint dwarf in the constellation of Pegasus. Pegasus~V (Peg~V) was initially identified in the public imaging data release of the DESI Legacy Imaging Surveys and confirmed with deep imaging from Gemini/GMOS-N. The colour-magnitude diagram shows a sparse red giant branch (RGB) population and a strong over-density of blue horizontal branch stars. We measure a distance to Peg~V of $D=692^{+33}_{-31}$~kpc, making it a distant satellite of Andromeda with $M_V=-6.3\pm0.2$ and a half-light radius of $r_{\rm half}=89\pm41$~pc. It is located $\sim260$~kpc from Andromeda in the outskirts of its halo. The RGB is well-fit by a metal-poor isochrone with [Fe/H]$=-3.2$, suggesting it is very metal poor. This, combined with its blue horizontal branch could imply that it is a reionisation fossil. This is the first detection of an ultra-faint dwarf outside the deep Pan-Andromeda Archaeological Survey area, and points to a rich, faint satellite population in the outskirts of our nearest neighbour. \end{abstract}

\begin{keywords}
galaxies: individual -- galaxies: dwarf
\end{keywords}



\section{Introduction}

The last two decades have seen an explosion in the detection of faint dwarf galaxies in the Local Group \cite[e.g.][]{willman05a,willman05b,belokurov06a,belokurov07a,belokurov08a,mcconnachie08,martin09,bechtol15,koposov15,torrealba16a,torrealba18,torrealba19, mau20,cerny21a,cerny21b}. Wide-field imaging surveys such as the Sloan Digital Sky Survey \citep{abazajian09}, Pan-STARRS \citep{chambers16}, the Pan-Andromeda Archaeological Survey (PAndAS, \citealt{mcconnachie09}), the Dark Energy Survey (DES, \citealt{abbott18}) and the DECam Local Volume Exploration Survey (DELVE, \citealt{drlica21}) have uncovered a wealth of substructures orbiting the Milky Way (MW) and Andromeda (M31) galaxies. Despite the discovery of dozens of new dwarf satellites, it is complex to reconcile the numbers of satellites with theoretical predictions \cite[e.g.][]{tollerud08,koposov09,walsh09,bovill11b,bullock17,kim18}. 

\begin{figure*}
	\includegraphics[width=6.0in]{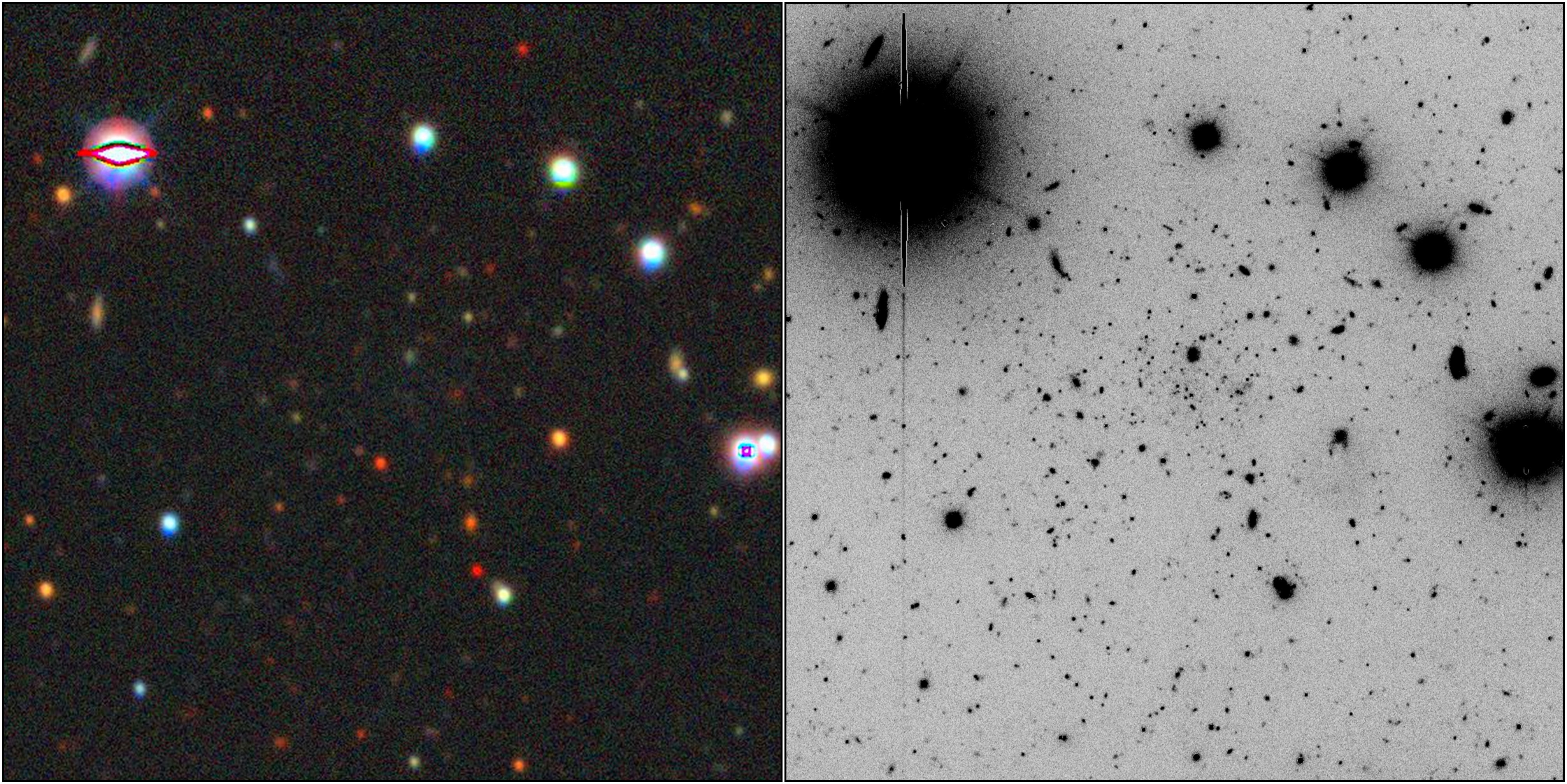}
    \caption{{\it Left panel}: Image of the dwarf galaxy Peg~V from the DESI Legacy Imaging Survey. {\it Right panel}: Gemini/GMOS-N combined $g-$ and $r-$band image of the galaxy obtained from Gemini follow-up observations (see Sec. 2). North is up, East is left. In both images, the field of view is 150$^{\prime\prime}\times150^{\prime\prime}$.}
    \label{fig:images}
\end{figure*}

At the brighter end ($L\gtrsim10^5\,{\rm L}_\odot$),  studies find that predictions and observations are in good agreement \cite[e.g.][]{sawala16,read19,kim18,engler21}. As such, it is likely that the remaining gap will be resolved as we push to ever-lower luminosities with new surveys, such as the Vera C. Rubin Legacy Survey of Space and Time \citep{tollerud08,bovill11a}. The faintest galaxies detected thus far -- the ultra-faint dwarfs (UFDs) with $M_V>-7.7$ \citep{bullock17} -- have only been detected nearby or in deep, focused surveys \cite[e.g.][]{belokurov07a,geha09,richardson11}. However, based on $\Lambda$ cold dark matter ($\Lambda$CDM) cosmology, they are expected to populate the full halos of the MW and M31, and the wider field \cite[e.g.][]{bovill11a,fattahi20a}. In M31, detection of the faintest galaxies has been limited to the central 150~kpc probed by PAndAS, which has discovered dwarf galaxies down to luminosities of $L\sim2\times10^4\,{\rm L_\odot}$ ($M_V\sim-6$, e.g. \citealt{mcconnachie09,richardson11,martin13c}). Work by \citet{martin16} demonstrated that there are likely many more faint dwarfs within this region, lurking just below the detection limit of the survey, and discoveries of brighter dwarfs at much greater distances ($\gtrsim200$~kpc) in shallower surveys \citep{martin13a,martin13b} suggest that there is a wealth of galaxies to be found out to the virial radius of the system.

These faintest dwarfs provide tight constraints on reionisation and stellar feedback physics as their gas reservoirs can more easily escape their low mass potentials. Deep imaging studies with the Hubble Space Telescope have shown that UFDs tend to have their star formation quenched early \cite[e.g.][]{brown14,sacchi21}, representing the relics of the very first galaxies \citep{bovill09}. They are also extremely dark matter dominated \cite[e.g.][]{simon07,martin07,geha09,simon19} and so their detection and further study may allow us insight into the nature of the dark matter particle.

With the value of finding new dwarfs to address these problems in mind, we have been conducting a visual inspection of imaging data from the DESI Legacy Imaging Survey \citep{Dey2019} in the vicinity of M31 and Triangulum (M33) to search for ultra-faint companions to both systems. We have already identified a potential UFD satellite of M33 -- Pisces VII/Triangulum III \citep{martinezdelgado22}. In this paper, we report on a newly discovered UFD satellite of M31 in the constellation of Pegasus -- Pegasus (Peg) V. In \S~\ref{sec:obs} we discuss the DESI Legacy Survey imaging, plus our deep follow-up observations with Gemini/GMOS-N, then in \S~\ref{sec:results} we present the properties of, and distance to, Peg~V. Finally we discuss the significance of our findings in \S~\ref{sec:discussion}.

\begin{figure*}
	\includegraphics[width=7.in]{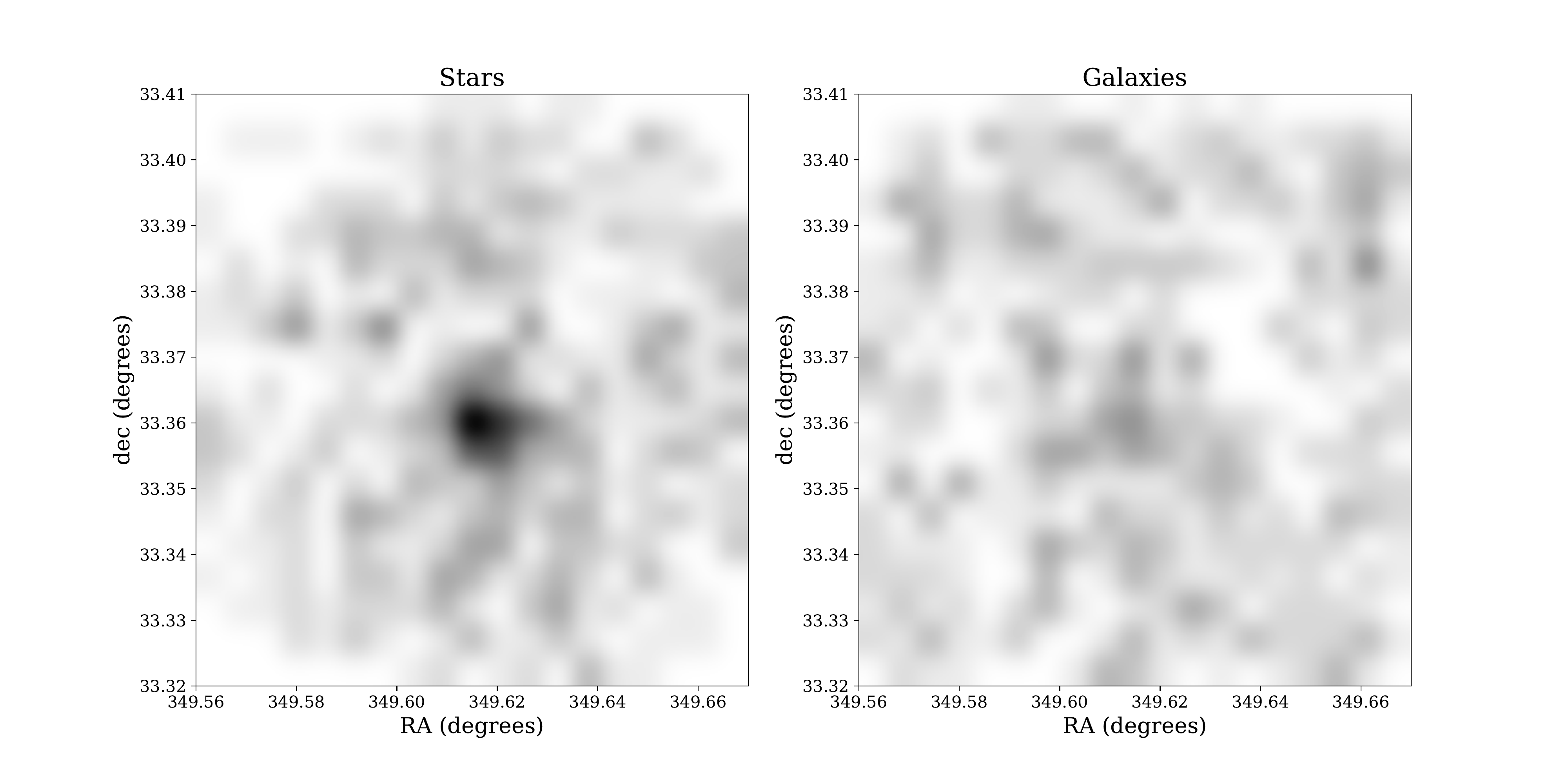}
    \caption{Filtered maps of all sources considered stars  ({\it left}) vs galaxies ({\it right}) within our GMOS-N field of view. A clear over-density is seen in the stellar map which is not reproduced in the galaxies panel. }
    \label{fig:map}
\end{figure*}

\section{Identification and Gemini Observations}
\label{sec:obs}

Peg V was identified as a partially resolved over-density in the DESI Legacy Imaging Survey (see left panel of Fig.~\ref{fig:images}) by amateur astronomer Giuseppe Donatiello. Given its position on-sky, it was a good candidate for an ultra-faint M31 or M33 satellite (projected distance of $\sim240$~kpc and 400~kpc from each respectively). We were awarded directors discretionary time (proposal ID GN-2021B-DD-106) to perform deep imaging with Gemini/GMOS-N with the aim of resolving its stellar populations down to its horizontal branch (HB). These observations were carried out on the nights of 2021-11-01 and 2021-11-25 and  employed the $g-$ and $r-$band filters. The total imaging time was 2250~s in $g-$band (split into $5\times450\,{\rm s}$ exposures) and 1500~s in $r-$band (split into $5\times300\,{\rm s}$ exposures). The images were pre-processed using the Gemini {\sc Dragons} pipeline which performs standard bias, flat field and cosmic ray corrections to the images before producing a final stacked image (which can be seen in the right hand panel of Fig.~\ref{fig:images}). The stacked images were then reduced using DAOPHOT/ALLFRAME \citep{stetson87,stetson94} in largely the same manner as \citet{monelli10b} and \citet{martinezdelgado22}. Briefly, we search for stellar sources on each stacked image and then perform aperture photometry, PSF derivation and PSF photometry with ALLSTAR.
The resulting list of stars is then passed to ALLFRAME to construct individual catalogues with better determined position and instrumental
magnitude of the input sources and provides the final photometry. We perform the photometric calibration using local standard stars from the Pan-STARRS1 $3 \pi$ survey
\citep{chambers16}. The mean magnitudes were
calibrated with a linear relation for the $g-$band, and a zero point for the $r-$band.
We extinction correct the data using the reddening maps from \citet{schlegeldust}, re-calibrated by \citet{schlafly11}.

Star-galaxy separation was performed using the sharpness parameter. In the left panel of Fig.~\ref{fig:map} we show the distribution of stars across the GMOS field of view, with the right panel showing the galaxy distribution. A clear over-density can be seen in the centre of the stellar map. In in the left panel of Fig.~\ref{fig:cmds} we show a colour-magnitude diagram (CMD) for all stars within 0.8 arcmin of this over-density compared to an equal area annulus beyond 1.6 arcmin in the centre panel. An old (12.5 Gyr), metal-poor ([Fe/H]$=-3.2$) BaSTI isochrone \citep{hidalgo18}\footnote{\href{http://basti-iac.oa-abruzzo.inaf.it/isocs.html}{http://basti-iac.oa-abruzzo.inaf.it/isocs.html}} is overlaid, shifted to a distance of $\sim700$~kpc to highlight what we assume is a red-giant branch (RGB) and horizontal branch (HB) feature.

\section{Properties of Pegasus~V}
\label{sec:results}


We determine the structural parameters for Peg~V using the same Markov-Chain Monte Carlo (MCMC) inference technique as presented in \citet{martinezdelgado22}. We defer to that work for a detailed discussion, and briefly outline this process below. The method is based on that of \citet{martin16} and uses {\sc emcee} \citep{emcee} to sample the posterior. We include all stars which fall within a broad colour-magnitude cut of $r_0<25.5$ and $-0.5<(g-r)_0<0.9$, giving 79 stars total. We then assume these stars comprise both members of the dwarf galaxy, and a foreground/background population. We assume the radial density profile, $\rho_{\rm dwarf}$ of the dwarf can be modelled as an exponential profile, with an ellipticity of $\epsilon = 1- (b/a)$ and $r_{\rm half}$ is the half-light radius. $N^*$ is the number of likely member stars inside the CMD selection box associated to the dwarf.





We assume the fore/background contamination, $\Sigma_b$, is constant across the field and determine this by subtracting the dwarf galaxy stars (which we calculate by integrating the radial density profile) from the total number of potential members identified, $n$. 


We combine these assumptions into a likelihood function which is then used in the MCMC analysis:

\begin{equation}
    \rho_{\rm model}(r) = \rho_{\rm dwarf}(r) + \Sigma_{b} .
    \label{eq:total profile}
\end{equation}

\begin{figure}
	\includegraphics[width=\columnwidth]{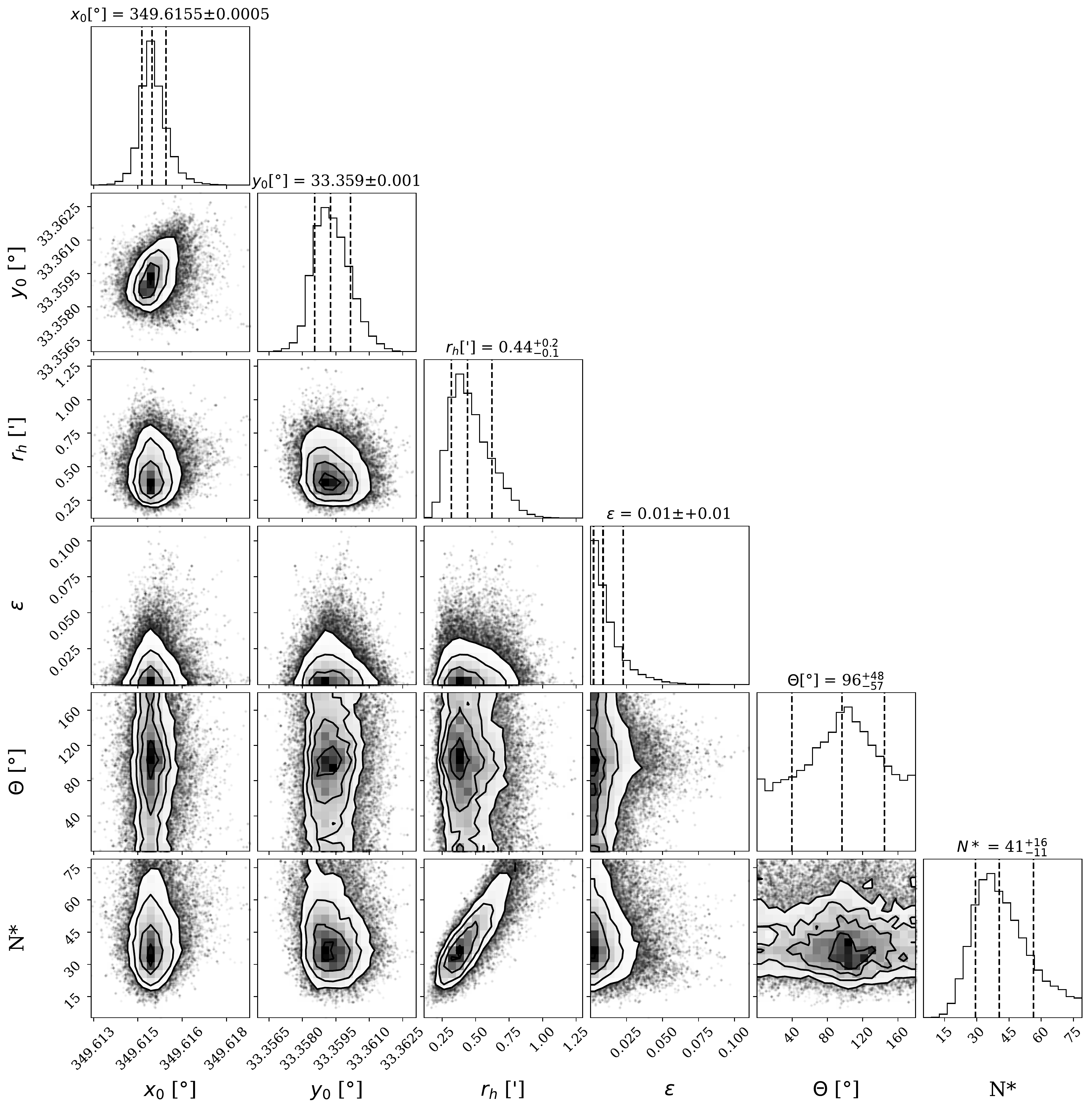}
	\caption{A corner plot showing the 2D and marginalised posteriors for the structural parameters of Peg~V. We show the central coordinates, $x_{0}$ and $y_{0}$, the half-light radius, $r_{\rm half}$, ellipticity, $\epsilon$, position angle, $\theta$ and the number of stars belonging to Peg~V, $N_*$ . The dashed lines represent the median value and 1$\sigma$ uncertainties.}
    \label{fig:corner}
\end{figure}

 We use broad, flat uniform priors for all parameters. The {\sc emcee} routine used 100 walkers, over a total of 10,000 iterations with a burn in of 9,250. These numbers were determined based on a visual inspection of the traces. We show the final corner plot in Fig. \ref{fig:corner} and summarise the resulting structural parameters and their 68\% confidence intervals in Table \ref{tab:structural params}. We find a position for Peg~V of $\alpha=$23:18:27.8$\pm0.1$, $\delta=$33:21:32$\pm3$, a half-light radius of $r_{\rm half}=0.44^{+0.2}_{-0.1}$ arcmin, and do not resolve an ellipticity.


\begin{figure*}
	\includegraphics[width=7.in]{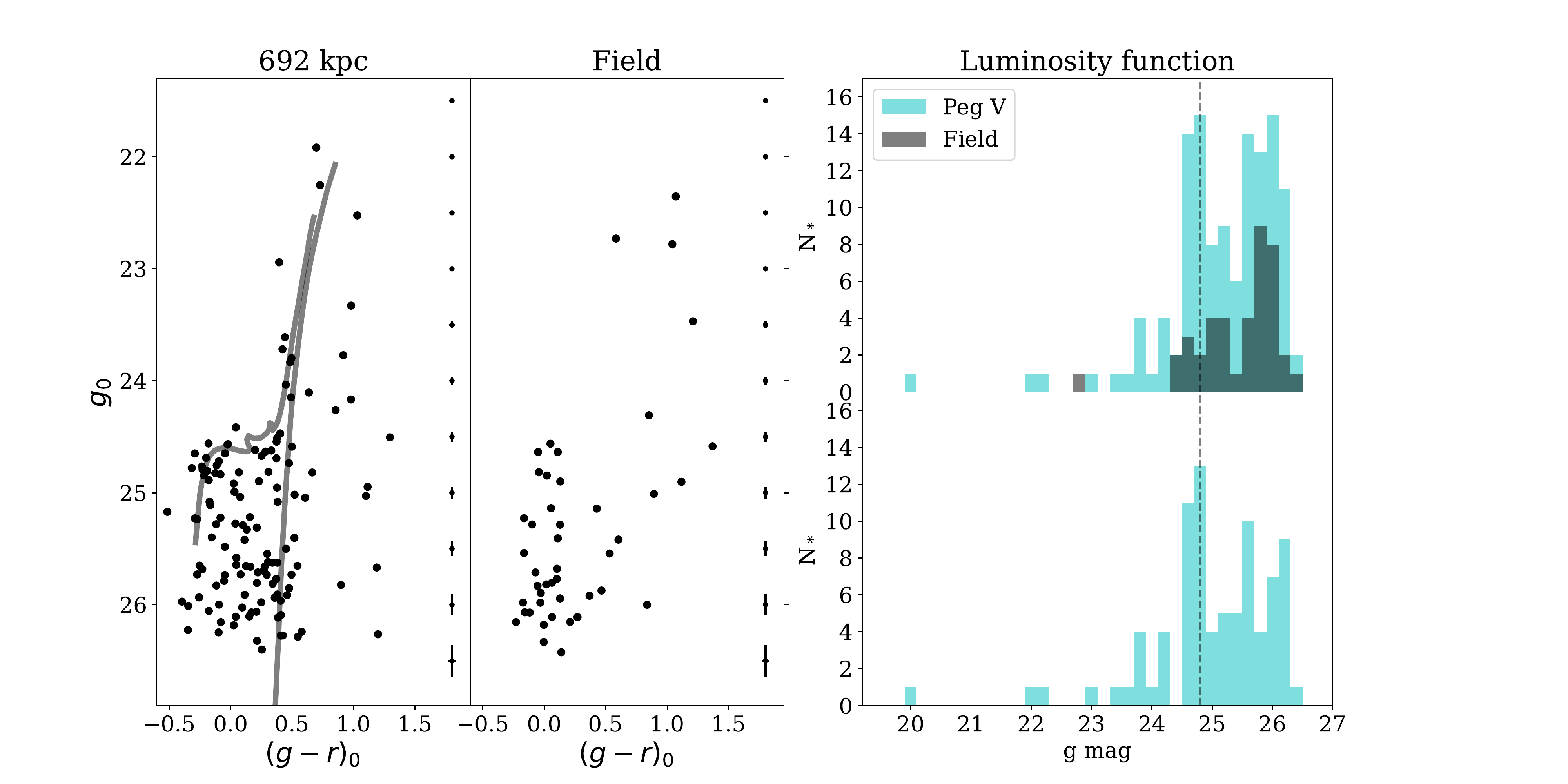}
    \caption{In the first two panels, we show the CMD for all stars within $2r_{\rm half}$ of Peg~V (left) compared to the stellar populations in an equal area annulus beyond the dwarf to demonstrate the field contamination (right). We overlay a metal poor ([Fe/H]$=-3.2$) isochrone with age 12.5 Gyr shifted to 692~kpc (centre), representing the distance measured from the HB. Finally, we show the luminosity function in the right most panels. The top shows all stars within $2r_{\rm half}$ (cyan) and the field population (grey), and the lower panel shows the background corrected case. The dashed grey line shows the position of the blue HB feature.}
    \label{fig:cmds}
\end{figure*}

A visual inspection of the CMD for Peg~V presents us with a sparse RGB and a well-populated HB feature. Given its projected distance from Andromeda of $D_{\rm proj}\sim245$~kpc, the most likely scenario is that Peg~V is an M31 satellite. 
To determine its distance, we use two techniques that focus on the RGB and HB populations. For the RGB, we employ the same Bayesian tip of the red giant branch (TRGB) approach as in \citet{tollerud16a} and \citet{martinezdelgado22}, which we briefly summarise below. We model the RGB and background population simultaneously as a broken power law with slopes of $\alpha$ and $\beta$ in luminosity respectively. We assume a fraction $f$ of stars in the background population (such that the fraction on the RGB is $1-f$) with the break located at the TRGB, $m_{\rm TRGB}$. 

We use Bayesian inference to determine posterior probabilities for $m_{\rm TRGB}$, using uniform priors for all parameters, with U(0, 1) for $f$ and U(0, 2) for $\alpha$ and $\beta$. For $m_{\rm TRGB}$ we use a broad prior of U(17, 22.5). We use the same colour selection as for the structural parameters of ($-0.5<g-r<0.9$), but only include stars with $r_0<24$ to isolate the RGB. We use only stars within 2$r_{\rm half}$, as if we include stars beyond this, the model has trouble distinguishing between background and dwarf (given the sparsity of RGB stars). We numerically integrate this luminosity function over a grid of $m$ using the trapezoid rule for each set of parameters, and take the total likelihood as the sum of the logarithm of this over all stars.

We use {\sc emcee} to sample the posterior distribution for the model parameters and find a solution of $m_{\rm TRGB}=21.2^{+1.0}_{-1.8}$, corresponding to a distance of $D=682^{+391}_{-355}$~kpc. This fit is poorly constrained owing to the small number of sources on the RGB.

A more precise distance can be determined by assuming the stars at $g_0\sim24.5$ constitute the HB. In the right-hand panels of Fig.~\ref{fig:cmds} we show the luminosity function in the $g-$band for Peg~V. There is a pronounced bump at $g_0\sim24.8\pm0.1$ which we assume is the HB. With $M_{g,{\rm HB}}=0.6$ \citep{irwin07}, we measure a distance of $D=692^{+33}_{-31}$~kpc, entirely consistent with the RGB approach but with far smaller uncertainties.

 In panel 1 of Fig.~\ref{fig:cmds} we show our isochrones shifted to this distance. We see that the HB and RGB features for the 12.5 Gyr isochrone perfectly overlap with the stellar populations of Peg~V, suggestsing that our candidate is an extremely metal poor M31 satellite.

 \begin{table}
	\centering
	\caption{The final structural and photometric properties for Peg~V.}
	\label{tab:structural params}
	\begin{tabular}{ll}
		\hline
		Property & Value\\
		\hline
		RA & $23^{h}$ $18^{m}$ $27.8^{s}$ $\pm$ ${0.1^{s}}$\\
		Dec & 33\textdegree $ 21^{\prime}$ $ 32^{\prime\prime}$ $\pm$ $3^{\prime\prime}$\\
        \vspace{0.1cm}
		$r_{\rm half}$ (arcmin) & 0.4$^{+0.2}_{-0.1}$\\
        \vspace{0.1cm}
		$D$ (kpc) & $692^{+33}_{-31}$\\
		$D_{\rm M31}$ (kpc) & $242^{+12}_{-11}$\\
		$r_{\rm half}$ (pc) & $87^{+40}_{-54}$\\  
		M$_V$ & $-6.3\pm0.2$\\
		$\mu_0$ (mag arcsec$^{-2}$) & $26.3\pm0.3$\\
		$L$  ($L_\odot$) & $2.8^{+0.6}_{-0.4}\times 10^4$\\
		$\epsilon$ & 0.01 $\pm0.01$\\
		$\theta$ (\textdegree) & 96 $^{{+{48}}}_{{-{57}}}$\\
		N$^*$ & 41 $^{{+{16}}}_{{-{11}}}$\\
		\hline
	\end{tabular}
\end{table}
Armed with a distance, we follow the approach of \citet{martin16} and \citet{martinezdelgado22} to measure the luminosity of Peg~V. Using theoretical luminosity functions generated  the PARSEC stellar models \citep{marigo08,girardi10}, we define a probability distribution function (PDF) which describes the expected number of RGB stars per magnitude bin. To generate the PDF, we use an old (12 Gyr), metal-poor ([Fe/H]$=-2.5$), $\alpha$ enhanced ($[\alpha/{\rm Fe}]=+0.4$) isochrone and a Kroupa IMF. We then randomly sample from this luminosity function to reproduce our observed dwarf which is simplified as having $N_*$ stars (where $N_*$ is taken from our MCMC analysis for the structural properties) above our magnitude cut of $r_0<25.5$. We record the sampled magnitude for all stars above and below this cut then convert these magnitudes into a luminosity using the distance measured from our TRGB analysis above. Once $N_*$ is reached, we sum all stars that were randomly sampled to get the total luminosity of Peg~V. We repeat this procedure 1000 times to measure the average luminosity plus a statistical uncertainty for each observed magnitude band. We then convert these to the $V$ band to obtain  $M_V=-6.3\pm0.2$ for a distance of $D=692^{+33}_{-31}$~kpc. 

\section{Discussion \& Conclusions}
\label{sec:discussion}

We have discovered a new Local Group ultra-faint dwarf in the constellation of Pegasus and characterised it using deep Gemini/GMOS-N imaging. Peg~V is a far-flung M31 satellite with a strong blue HB, $M_V=-6.3\pm0.2$, located 242~kpc from the centre of the system. This would place it close to the virial radius of M31 (estimated to be between $\sim250-300$~kpc, e.g. \citealt{patel18a,kafle18,blana18}). It would also be the faintest M31 satellite detected outside of the area covered by the deep PAndAS Survey. 

The strong blue HB of Peg~V makes it somewhat unique when compared with the known satellite population of Andromeda. M31 dwarf satellites typically possess far redder HBs (e.g. \citealt{dacosta96, dacosta00,dacosta02}) with only a handful showing a significant blue component. \citet{martin17} present homogeneous single orbit HST observations for 20 M31 satellite galaxies which well-resolve their HBs. They assess the colour of their HB features by computing the ratio, $\eta$, of the number of blue HB stars ($n_{\rm BHB}$) to the total number of HB stars ($n_{\rm BHB}+n_{\rm RHB}$). They find only two examples of dwarfs with relatively high fractions of blue HB stars, giving $\eta\sim0.4$ (Andromeda XI and XVII). If we repeat their analysis for Peg~V, we find $\eta=0.5\pm 0.1$, setting it apart from the rest of the system. As blue HB stars are typically considered to trace ancient stellar populations, this could imply Peg~V was quenched very early compared to the other M31 dwarfs, making it a good candidate for a reionisation fossil \citep{bovill09}. It is also more similar to the UFDs of the MW, all of which appear to have been quenched $\gtrsim10$~Gyr ago \citep{brown14,weisz14,sacchi21} and possess blue HB features.

The RGB stars of Peg~V are also extremely blue, and are well described by the most metal-poor BaSTI isochrone (see Fig.~\ref{fig:cmds}), suggesting it is a very metal-poor dwarf galaxy. Combined with its blue HB, this may imply it underwent very little chemical enrichment before having its star formation rapidly quenched by reionisation. If true, further study of its chemical properties may allow us to place constraints on the earliest epochs of star formation.

With deep space-based imaging we could further probe the star formation history of Peg~V, and learn if it is a candidate for a fossil galaxy quenched by reionisation. It also has a few bright RGB stars which can be targeted with 8-10m telescopes to measure their velocities and chemistry. Given its remote position in the halo, kinematics may also allow constraints of its orbit around M31 and the likelihood of previous interactions with its host that may also have quenched its star formation.

 The discovery of Peg~V in the DESI Legacy Imaging Survey bodes well for future discoveries of ultra-faint dwarfs in M31 and the wider Local Group with current and future surveys. It paints a rosy picture for simultaneously solving both the missing satellite and field galaxy problems in our cosmic backyard. Our systematic visual inspection of the DESI images successfully detects partially resolved ultra-faint dwarf satellites in the M31/M33 system, which might be overlooked in the automatic detection of over-densities in stellar density maps (e.g. DELVE). Thus, both approaches are extremely complementary.

\section*{Acknowledgements}

DMD acknowledges financial support from the Talentia Senior Program (through the incentive ASE-136) from Secretar\'\i a General de  Universidades, Investigaci\'{o}n y Tecnolog\'\i a, de la Junta de Andaluc\'\i a. DMD acknowledge funding from the State Agency for Research of the Spanish MCIU through the ``Center of Excellence Severo Ochoa" award to the Instituto de Astrof{\'i}sica de Andaluc{\'i}a (SEV-2017-0709) and project (PDI2020-114581GB-C21/ AEI / 10.13039/501100011033)

Based on observations obtained at the international Gemini Observatory, a program of NSF’s NOIRLab, which is managed by the Association of Universities for Research in Astronomy (AURA) under a cooperative agreement with the National Science Foundation. on behalf of the Gemini Observatory partnership: the National Science Foundation (United States), National Research Council (Canada), Agencia Nacional de Investigaci\'{o}n y Desarrollo (Chile), Ministerio de Ciencia, Tecnolog\'{i}a e Innovaci\'{o}n (Argentina), Minist\'{e}rio da Ci\^{e}ncia, Tecnologia, Inova\c{c}\~{o}es e Comunica\c{c}\~{o}es (Brazil), and Korea Astronomy and Space Science Institute (Republic of Korea).

Data processed using DRAGONS (Data Reduction for Astronomy from Gemini Observatory North and South).

This work was enabled by observations made from the Gemini North telescope, located within the Maunakea Science Reserve and adjacent to the summit of Maunakea. We are grateful for the privilege of observing the Universe from a place that is unique in both its astronomical quality and its cultural significance.

This project used public archival data from the {\it DESI Legacy Imaging Surveys} (DESI LIS). The Legacy Surveys consist of three individual and complementary projects: the Dark Energy Camera Legacy Survey (DECaLS; Proposal ID \#2014B-0404; PIs: David Schlegel and Arjun Dey), the Beijing-Arizona Sky Survey (BASS; NOAO Prop. ID \#2015A-0801; PIs: Zhou Xu and Xiaohui Fan), and the Mayall z-band Legacy Survey (MzLS; Prop. ID \#2016A-0453; PI: Arjun Dey). DECaLS, BASS and MzLS together include data obtained, respectively, at the Blanco telescope, Cerro Tololo Inter-American Observatory, NSF’s NOIRLab; the Bok telescope, Steward Observatory, University of Arizona; and the Mayall telescope, Kitt Peak National Observatory, NOIRLab. The Legacy Surveys project is honored to be permitted to conduct astronomical research on Iolkam Du’ag (Kitt Peak), a mountain with particular significance to the Tohono O’odham Nation.

NOIRLab is operated by the Association of Universities for Research in Astronomy (AURA) under a cooperative agreement with the National Science Foundation.

This project used data obtained with the Dark Energy Camera (DECam), which was constructed by the Dark Energy Survey (DES) collaboration. Funding for the DES Projects has been provided by the U.S. Department of Energy, the U.S. National Science Foundation, the Ministry of Science and Education of Spain, the Science and Technology Facilities Council of the United Kingdom, the Higher Education Funding Council for England, the National Center for Supercomputing Applications at the University of Illinois at Urbana-Champaign, the Kavli Institute of Cosmological Physics at the University of Chicago, Center for Cosmology and Astro-Particle Physics at the Ohio State University, the Mitchell Institute for Fundamental Physics and Astronomy at Texas A\&M University, Financiadora de Estudos e Projetos, Fundacao Carlos Chagas Filho de Amparo, Financiadora de Estudos e Projetos, Fundacao Carlos Chagas Filho de Amparo a Pesquisa do Estado do Rio de Janeiro, Conselho Nacional de Desenvolvimento Cientifico e Tecnologico and the Ministerio da Ciencia, Tecnologia e Inovacao, the Deutsche Forschungsgemeinschaft and the Collaborating Institutions in the Dark Energy Survey. The Collaborating Institutions are Argonne National Laboratory, the University of California at Santa Cruz, the University of Cambridge, Centro de Investigaciones Energeticas, Medioambientales y Tecnologicas-Madrid, the University of Chicago, University College London, the DES-Brazil Consortium, the University of Edinburgh, the Eidgenossische Technische Hochschule (ETH) Zurich, Fermi National Accelerator Laboratory, the University of Illinois at Urbana-Champaign, the Institut de Ciencies de l’Espai (IEEC/CSIC), the Institut de Fisica d’Altes Energies, Lawrence Berkeley National Laboratory, the Ludwig Maximilians Universitat Munchen and the associated Excellence Cluster Universe, the University of Michigan, NSF’s NOIRLab, the University of Nottingham, the Ohio State University, the University of Pennsylvania, the University of Portsmouth, SLAC National Accelerator Laboratory, Stanford University, the University of Sussex, and Texas A\&M University.

BASS is a key project of the Telescope Access Program (TAP), which has been funded by the National Astronomical Observatories of China, the Chinese Academy of Sciences (the Strategic Priority Research Program “The Emergence of Cosmological Structures” Grant \# XDB09000000), and the Special Fund for Astronomy from the Ministry of Finance. The BASS is also supported by the External Cooperation Program of Chinese Academy of Sciences (Grant \# 114A11KYSB20160057), and Chinese National Natural Science Foundation (Grant \# 11433005).

The Legacy Survey team makes use of data products from the Near-Earth Object Wide-field Infrared Survey Explorer (NEOWISE), which is a project of the Jet Propulsion Laboratory/California Institute of Technology. NEOWISE is funded by the National Aeronautics and Space Administration.

The Legacy Surveys imaging of the DESI footprint is supported by the Director, Office of Science, Office of High Energy Physics of the U.S. Department of Energy under Contract No. DE-AC02-05CH1123, by the National Energy Research Scientific Computing Center, a DOE Office of Science User Facility under the same contract; and by the U.S. National Science Foundation, Division of Astronomical Sciences under Contract No. AST-0950945 to NOAO.

\section*{Data Availability}

The DESI Legacy Imaging data are all publicly available at \url{www.legacysurvey.org}. The Gemini/GMOS-N images are hosted at \url{archive.gemini.edu/searchform}. They are associated with program ID: GN-2021B-DD-106 and made publicly available 6 months after acquisition. Reduced photometry can be obtained from the lead author upon reasonable request.



\bibliographystyle{mnras}
\bibliography{PegasusV} 

\begin{thebibliography}{}
\makeatletter
\relax
\def\mn@urlcharsother{\let\do\@makeother \do\$\do\&\do\#\do\^\do\_\do\%\do\~}
\def\mn@doi{\begingroup\mn@urlcharsother \@ifnextchar [ {\mn@doi@}
  {\mn@doi@[]}}
\def\mn@doi@[#1]#2{\def\@tempa{#1}\ifx\@tempa\@empty \href
  {http://dx.doi.org/#2} {doi:#2}\else \href {http://dx.doi.org/#2} {#1}\fi
  \endgroup}
\def\mn@eprint#1#2{\mn@eprint@#1:#2::\@nil}
\def\mn@eprint@arXiv#1{\href {http://arxiv.org/abs/#1} {{\tt arXiv:#1}}}
\def\mn@eprint@dblp#1{\href {http://dblp.uni-trier.de/rec/bibtex/#1.xml}
  {dblp:#1}}
\def\mn@eprint@#1:#2:#3:#4\@nil{\def\@tempa {#1}\def\@tempb {#2}\def\@tempc
  {#3}\ifx \@tempc \@empty \let \@tempc \@tempb \let \@tempb \@tempa \fi \ifx
  \@tempb \@empty \def\@tempb {arXiv}\fi \@ifundefined
  {mn@eprint@\@tempb}{\@tempb:\@tempc}{\expandafter \expandafter \csname
  mn@eprint@\@tempb\endcsname \expandafter{\@tempc}}}

\bibitem[\protect\citeauthoryear{{Abazajian} et~al.,}{{Abazajian}
  et~al.}{2009}]{abazajian09}
{Abazajian} K.~N.,  et~al., 2009, \mn@doi [\apjs]
  {10.1088/0067-0049/182/2/543}, \href
  {https://ui.adsabs.harvard.edu/abs/2009ApJS..182..543A} {182, 543}

\bibitem[\protect\citeauthoryear{{Abbott} et~al.,}{{Abbott}
  et~al.}{2018}]{abbott18}
{Abbott} T.~M.~C.,  et~al., 2018, \mn@doi [\apjs] {10.3847/1538-4365/aae9f0},
  \href {https://ui.adsabs.harvard.edu/abs/2018ApJS..239...18A} {239, 18}

\bibitem[\protect\citeauthoryear{{Bechtol} et~al.,}{{Bechtol}
  et~al.}{2015}]{bechtol15}
{Bechtol} K.,  et~al., 2015, \mn@doi [\apj] {10.1088/0004-637X/807/1/50}, \href
  {https://ui.adsabs.harvard.edu/abs/2015ApJ...807...50B} {807, 50}

\bibitem[\protect\citeauthoryear{{Belokurov} et~al.,}{{Belokurov}
  et~al.}{2006}]{belokurov06a}
{Belokurov} V.,  et~al., 2006, \mn@doi [\apjl] {10.1086/507324}, \href
  {https://ui.adsabs.harvard.edu/abs/2006ApJ...647L.111B} {647, L111}

\bibitem[\protect\citeauthoryear{{Belokurov} et~al.,}{{Belokurov}
  et~al.}{2007}]{belokurov07a}
{Belokurov} V.,  et~al., 2007, \mn@doi [\apj] {10.1086/509718}, \href
  {https://ui.adsabs.harvard.edu/abs/2007ApJ...654..897B} {654, 897}

\bibitem[\protect\citeauthoryear{{Belokurov} et~al.,}{{Belokurov}
  et~al.}{2008}]{belokurov08a}
{Belokurov} V.,  et~al., 2008, \mn@doi [\apjl] {10.1086/592962}, \href
  {https://ui.adsabs.harvard.edu/abs/2008ApJ...686L..83B} {686, L83}

\bibitem[\protect\citeauthoryear{{Bla{\~n}a D{\'\i}az} et~al.,}{{Bla{\~n}a
  D{\'\i}az} et~al.}{2018}]{blana18}
{Bla{\~n}a D{\'\i}az} M.,  et~al., 2018, \mn@doi [\mnras]
  {10.1093/mnras/sty2311}, \href
  {https://ui.adsabs.harvard.edu/abs/2018MNRAS.481.3210B} {481, 3210}

\bibitem[\protect\citeauthoryear{{Bovill} \& {Ricotti}}{{Bovill} \&
  {Ricotti}}{2009}]{bovill09}
{Bovill} M.~S.,  {Ricotti} M.,  2009, \mn@doi [\apj]
  {10.1088/0004-637X/693/2/1859}, \href
  {https://ui.adsabs.harvard.edu/abs/2009ApJ...693.1859B} {693, 1859}

\bibitem[\protect\citeauthoryear{{Bovill} \& {Ricotti}}{{Bovill} \&
  {Ricotti}}{2011a}]{bovill11a}
{Bovill} M.~S.,  {Ricotti} M.,  2011a, \mn@doi [\apj]
  {10.1088/0004-637X/741/1/17}, \href
  {https://ui.adsabs.harvard.edu/abs/2011ApJ...741...17B} {741, 17}

\bibitem[\protect\citeauthoryear{{Bovill} \& {Ricotti}}{{Bovill} \&
  {Ricotti}}{2011b}]{bovill11b}
{Bovill} M.~S.,  {Ricotti} M.,  2011b, \mn@doi [\apj]
  {10.1088/0004-637X/741/1/18}, \href
  {https://ui.adsabs.harvard.edu/abs/2011ApJ...741...18B} {741, 18}

\bibitem[\protect\citeauthoryear{{Brown} et~al.,}{{Brown}
  et~al.}{2014}]{brown14}
{Brown} T.~M.,  et~al., 2014, \mn@doi [\apj] {10.1088/0004-637X/796/2/91},
  \href {https://ui.adsabs.harvard.edu/abs/2014ApJ...796...91B} {796, 91}

\bibitem[\protect\citeauthoryear{{Bullock} \& {Boylan-Kolchin}}{{Bullock} \&
  {Boylan-Kolchin}}{2017}]{bullock17}
{Bullock} J.~S.,  {Boylan-Kolchin} M.,  2017, \mn@doi [\araa]
  {10.1146/annurev-astro-091916-055313}, \href
  {https://ui.adsabs.harvard.edu/abs/2017ARA&A..55..343B} {55, 343}

\bibitem[\protect\citeauthoryear{{Cerny} et~al.,}{{Cerny}
  et~al.}{2021a}]{cerny21a}
{Cerny} W.,  et~al., 2021a, \mn@doi [\apj] {10.3847/1538-4357/abe1af}, \href
  {https://ui.adsabs.harvard.edu/abs/2021ApJ...910...18C} {910, 18}

\bibitem[\protect\citeauthoryear{{Cerny} et~al.,}{{Cerny}
  et~al.}{2021b}]{cerny21b}
{Cerny} W.,  et~al., 2021b, \mn@doi [\apjl] {10.3847/2041-8213/ac2d9a}, \href
  {https://ui.adsabs.harvard.edu/abs/2021ApJ...920L..44C} {920, L44}

\bibitem[\protect\citeauthoryear{{Chambers} et~al.,}{{Chambers}
  et~al.}{2016}]{chambers16}
{Chambers} K.~C.,  et~al., 2016, arXiv e-prints, \href
  {https://ui.adsabs.harvard.edu/abs/2016arXiv161205560C} {p. arXiv:1612.05560}

\bibitem[\protect\citeauthoryear{{Da Costa}, {Armandroff}, {Caldwell}  \&
  {Seitzer}}{{Da Costa} et~al.}{1996}]{dacosta96}
{Da Costa} G.~S.,  {Armandroff} T.~E.,  {Caldwell} N.,   {Seitzer} P.,  1996,
  \mn@doi [\aj] {10.1086/118204}, \href
  {https://ui.adsabs.harvard.edu/abs/1996AJ....112.2576D} {112, 2576}

\bibitem[\protect\citeauthoryear{{Da Costa}, {Armandroff}, {Caldwell}  \&
  {Seitzer}}{{Da Costa} et~al.}{2000}]{dacosta00}
{Da Costa} G.~S.,  {Armandroff} T.~E.,  {Caldwell} N.,   {Seitzer} P.,  2000,
  \mn@doi [\aj] {10.1086/301223}, \href
  {https://ui.adsabs.harvard.edu/abs/2000AJ....119..705D} {119, 705}

\bibitem[\protect\citeauthoryear{{Da Costa}, {Armandroff}  \& {Caldwell}}{{Da
  Costa} et~al.}{2002}]{dacosta02}
{Da Costa} G.~S.,  {Armandroff} T.~E.,   {Caldwell} N.,  2002, \mn@doi [\aj]
  {10.1086/340965}, \href
  {https://ui.adsabs.harvard.edu/abs/2002AJ....124..332D} {124, 332}

\bibitem[\protect\citeauthoryear{{Dey} et~al.,}{{Dey} et~al.}{2019}]{Dey2019}
{Dey} A.,  et~al., 2019, \mn@doi [\aj] {10.3847/1538-3881/ab089d}, \href
  {https://ui.adsabs.harvard.edu/abs/2019AJ....157..168D} {157, 168}

\bibitem[\protect\citeauthoryear{{Drlica-Wagner} et~al.,}{{Drlica-Wagner}
  et~al.}{2021}]{drlica21}
{Drlica-Wagner} A.,  et~al., 2021, \mn@doi [\apjs] {10.3847/1538-4365/ac079d},
  \href {https://ui.adsabs.harvard.edu/abs/2021ApJS..256....2D} {256, 2}

\bibitem[\protect\citeauthoryear{{Engler} et~al.,}{{Engler}
  et~al.}{2021}]{engler21}
{Engler} C.,  et~al., 2021, \mn@doi [\mnras] {10.1093/mnras/stab2437}, \href
  {https://ui.adsabs.harvard.edu/abs/2021MNRAS.507.4211E} {507, 4211}

\bibitem[\protect\citeauthoryear{{Fattahi}, {Navarro}  \& {Frenk}}{{Fattahi}
  et~al.}{2020}]{fattahi20a}
{Fattahi} A.,  {Navarro} J.~F.,   {Frenk} C.~S.,  2020, \mn@doi [\mnras]
  {10.1093/mnras/staa375}, \href
  {https://ui.adsabs.harvard.edu/abs/2020MNRAS.493.2596F} {493, 2596}

\bibitem[\protect\citeauthoryear{{Foreman-Mackey}, {Hogg}, {Lang}  \&
  {Goodman}}{{Foreman-Mackey} et~al.}{2013}]{emcee}
{Foreman-Mackey} D.,  {Hogg} D.~W.,  {Lang} D.,   {Goodman} J.,  2013, \mn@doi
  [\pasp] {10.1086/670067}, \href
  {https://ui.adsabs.harvard.edu/abs/2013PASP..125..306F} {125, 306}

\bibitem[\protect\citeauthoryear{{Geha}, {Willman}, {Simon}, {Strigari},
  {Kirby}, {Law}  \& {Strader}}{{Geha} et~al.}{2009}]{geha09}
{Geha} M.,  {Willman} B.,  {Simon} J.~D.,  {Strigari} L.~E.,  {Kirby} E.~N.,
  {Law} D.~R.,   {Strader} J.,  2009, \mn@doi [\apj]
  {10.1088/0004-637X/692/2/1464}, \href
  {https://ui.adsabs.harvard.edu/abs/2009ApJ...692.1464G} {692, 1464}

\bibitem[\protect\citeauthoryear{{Girardi} et~al.,}{{Girardi}
  et~al.}{2010}]{girardi10}
{Girardi} L.,  et~al., 2010, \mn@doi [\apj] {10.1088/0004-637X/724/2/1030},
  \href {https://ui.adsabs.harvard.edu/abs/2010ApJ...724.1030G} {724, 1030}

\bibitem[\protect\citeauthoryear{{Hidalgo} et~al.,}{{Hidalgo}
  et~al.}{2018}]{hidalgo18}
{Hidalgo} S.~L.,  et~al., 2018, \mn@doi [\apj] {10.3847/1538-4357/aab158},
  \href {https://ui.adsabs.harvard.edu/abs/2018ApJ...856..125H} {856, 125}

\bibitem[\protect\citeauthoryear{{Irwin} et~al.,}{{Irwin}
  et~al.}{2007}]{irwin07}
{Irwin} M.~J.,  et~al., 2007, \mn@doi [\apjl] {10.1086/512183}, \href
  {https://ui.adsabs.harvard.edu/abs/2007ApJ...656L..13I} {656, L13}

\bibitem[\protect\citeauthoryear{{Kafle}, {Sharma}, {Lewis}, {Robotham}  \&
  {Driver}}{{Kafle} et~al.}{2018}]{kafle18}
{Kafle} P.~R.,  {Sharma} S.,  {Lewis} G.~F.,  {Robotham} A. S.~G.,   {Driver}
  S.~P.,  2018, \mn@doi [\mnras] {10.1093/mnras/sty082}, \href
  {https://ui.adsabs.harvard.edu/abs/2018MNRAS.475.4043K} {475, 4043}

\bibitem[\protect\citeauthoryear{{Kim}, {Peter}  \& {Hargis}}{{Kim}
  et~al.}{2018}]{kim18}
{Kim} S.~Y.,  {Peter} A. H.~G.,   {Hargis} J.~R.,  2018, \mn@doi [\prl]
  {10.1103/PhysRevLett.121.211302}, \href
  {https://ui.adsabs.harvard.edu/abs/2018PhRvL.121u1302K} {121, 211302}

\bibitem[\protect\citeauthoryear{{Koposov}, {Yoo}, {Rix}, {Weinberg},
  {Macci{\`o}}  \& {Escud{\'e}}}{{Koposov} et~al.}{2009}]{koposov09}
{Koposov} S.~E.,  {Yoo} J.,  {Rix} H.-W.,  {Weinberg} D.~H.,  {Macci{\`o}}
  A.~V.,   {Escud{\'e}} J.~M.,  2009, \mn@doi [\apj]
  {10.1088/0004-637X/696/2/2179}, \href
  {https://ui.adsabs.harvard.edu/abs/2009ApJ...696.2179K} {696, 2179}

\bibitem[\protect\citeauthoryear{{Koposov}, {Belokurov}, {Torrealba}  \&
  {Evans}}{{Koposov} et~al.}{2015}]{koposov15}
{Koposov} S.~E.,  {Belokurov} V.,  {Torrealba} G.,   {Evans} N.~W.,  2015,
  \mn@doi [\apj] {10.1088/0004-637X/805/2/130}, \href
  {https://ui.adsabs.harvard.edu/abs/2015ApJ...805..130K} {805, 130}

\bibitem[\protect\citeauthoryear{{Marigo}, {Girardi}, {Bressan}, {Groenewegen},
  {Silva}  \& {Granato}}{{Marigo} et~al.}{2008}]{marigo08}
{Marigo} P.,  {Girardi} L.,  {Bressan} A.,  {Groenewegen} M.~A.~T.,  {Silva}
  L.,   {Granato} G.~L.,  2008, \mn@doi [\aap] {10.1051/0004-6361:20078467},
  \href {https://ui.adsabs.harvard.edu/abs/2008A&A...482..883M} {482, 883}

\bibitem[\protect\citeauthoryear{{Martin}, {Ibata}, {Chapman}, {Irwin}  \&
  {Lewis}}{{Martin} et~al.}{2007}]{martin07}
{Martin} N.~F.,  {Ibata} R.~A.,  {Chapman} S.~C.,  {Irwin} M.,   {Lewis} G.~F.,
   2007, \mn@doi [\mnras] {10.1111/j.1365-2966.2007.12055.x}, \href
  {https://ui.adsabs.harvard.edu/abs/2007MNRAS.380..281M} {380, 281}

\bibitem[\protect\citeauthoryear{{Martin} et~al.,}{{Martin}
  et~al.}{2009}]{martin09}
{Martin} N.~F.,  et~al., 2009, \mn@doi [\apj] {10.1088/0004-637X/705/1/758},
  \href {https://ui.adsabs.harvard.edu/abs/2009ApJ...705..758M} {705, 758}

\bibitem[\protect\citeauthoryear{{Martin} et~al.,}{{Martin}
  et~al.}{2013a}]{martin13a}
{Martin} N.~F.,  et~al., 2013a, \mn@doi [\apj] {10.1088/0004-637X/772/1/15},
  \href {https://ui.adsabs.harvard.edu/abs/2013ApJ...772...15M} {772, 15}

\bibitem[\protect\citeauthoryear{{Martin}, {Ibata}, {McConnachie}, {Mackey},
  {Ferguson}, {Irwin}, {Lewis}  \& {Fardal}}{{Martin}
  et~al.}{2013b}]{martin13c}
{Martin} N.~F.,  {Ibata} R.~A.,  {McConnachie} A.~W.,  {Mackey} A.~D.,
  {Ferguson} A. M.~N.,  {Irwin} M.~J.,  {Lewis} G.~F.,   {Fardal} M.~A.,
  2013b, \mn@doi [\apj] {10.1088/0004-637X/776/2/80}, \href
  {https://ui.adsabs.harvard.edu/abs/2013ApJ...776...80M} {776, 80}

\bibitem[\protect\citeauthoryear{{Martin} et~al.,}{{Martin}
  et~al.}{2013c}]{martin13b}
{Martin} N.~F.,  et~al., 2013c, \mn@doi [\apjl] {10.1088/2041-8205/779/1/L10},
  \href {https://ui.adsabs.harvard.edu/abs/2013ApJ...779L..10M} {779, L10}

\bibitem[\protect\citeauthoryear{{Martin} et~al.,}{{Martin}
  et~al.}{2016}]{martin16}
{Martin} N.~F.,  et~al., 2016, \mn@doi [\apj] {10.3847/1538-4357/833/2/167},
  \href {https://ui.adsabs.harvard.edu/abs/2016ApJ...833..167M} {833, 167}

\bibitem[\protect\citeauthoryear{{Martin} et~al.,}{{Martin}
  et~al.}{2017}]{martin17}
{Martin} N.~F.,  et~al., 2017, \mn@doi [\apj] {10.3847/1538-4357/aa901a}, \href
  {https://ui.adsabs.harvard.edu/abs/2017ApJ...850...16M} {850, 16}

\bibitem[\protect\citeauthoryear{{Mart{\'\i}nez-Delgado}, {Karim}, {Charles},
  {Boschin}, {Monelli}, {Collins}, {Donatiello}  \&
  {Alfaro}}{{Mart{\'\i}nez-Delgado} et~al.}{2022}]{martinezdelgado22}
{Mart{\'\i}nez-Delgado} D.,  {Karim} N.,  {Charles} E. J.~E.,  {Boschin} W.,
  {Monelli} M.,  {Collins} M. L.~M.,  {Donatiello} G.,   {Alfaro} E.~J.,  2022,
  \mn@doi [\mnras] {10.1093/mnras/stab2797}, \href
  {https://ui.adsabs.harvard.edu/abs/2022MNRAS.509...16M} {509, 16}

\bibitem[\protect\citeauthoryear{{Mau} et~al.,}{{Mau} et~al.}{2020}]{mau20}
{Mau} S.,  et~al., 2020, \mn@doi [\apj] {10.3847/1538-4357/ab6c67}, \href
  {https://ui.adsabs.harvard.edu/abs/2020ApJ...890..136M} {890, 136}

\bibitem[\protect\citeauthoryear{{McConnachie} et~al.,}{{McConnachie}
  et~al.}{2008}]{mcconnachie08}
{McConnachie} A.~W.,  et~al., 2008, \mn@doi [\apj] {10.1086/591313}, \href
  {https://ui.adsabs.harvard.edu/abs/2008ApJ...688.1009M} {688, 1009}

\bibitem[\protect\citeauthoryear{{McConnachie} et~al.,}{{McConnachie}
  et~al.}{2009}]{mcconnachie09}
{McConnachie} A.~W.,  et~al., 2009, \mn@doi [\nat] {10.1038/nature08327}, \href
  {https://ui.adsabs.harvard.edu/abs/2009Natur.461...66M} {461, 66}

\bibitem[\protect\citeauthoryear{{Monelli} et~al.,}{{Monelli}
  et~al.}{2010}]{monelli10b}
{Monelli} M.,  et~al., 2010, \mn@doi [\apj] {10.1088/0004-637X/720/2/1225},
  \href {https://ui.adsabs.harvard.edu/abs/2010ApJ...720.1225M} {720, 1225}

\bibitem[\protect\citeauthoryear{{Patel}, {Besla}, {Mandel}  \& {Sohn}}{{Patel}
  et~al.}{2018}]{patel18a}
{Patel} E.,  {Besla} G.,  {Mandel} K.,   {Sohn} S.~T.,  2018, \mn@doi [\apj]
  {10.3847/1538-4357/aab78f}, \href
  {https://ui.adsabs.harvard.edu/abs/2018ApJ...857...78P} {857, 78}

\bibitem[\protect\citeauthoryear{{Read} \& {Erkal}}{{Read} \&
  {Erkal}}{2019}]{read19}
{Read} J.~I.,  {Erkal} D.,  2019, \mn@doi [\mnras] {10.1093/mnras/stz1320},
  \href {https://ui.adsabs.harvard.edu/abs/2019MNRAS.487.5799R} {487, 5799}

\bibitem[\protect\citeauthoryear{{Richardson} et~al.,}{{Richardson}
  et~al.}{2011}]{richardson11}
{Richardson} J.~C.,  et~al., 2011, \mn@doi [\apj] {10.1088/0004-637X/732/2/76},
  \href {https://ui.adsabs.harvard.edu/abs/2011ApJ...732...76R} {732, 76}

\bibitem[\protect\citeauthoryear{{Sacchi} et~al.,}{{Sacchi}
  et~al.}{2021}]{sacchi21}
{Sacchi} E.,  et~al., 2021, \mn@doi [\apjl] {10.3847/2041-8213/ac2aa3}, \href
  {https://ui.adsabs.harvard.edu/abs/2021ApJ...920L..19S} {920, L19}

\bibitem[\protect\citeauthoryear{{Sawala} et~al.,}{{Sawala}
  et~al.}{2016}]{sawala16}
{Sawala} T.,  et~al., 2016, \mn@doi [\mnras] {10.1093/mnras/stw145}, \href
  {https://ui.adsabs.harvard.edu/abs/2016MNRAS.457.1931S} {457, 1931}

\bibitem[\protect\citeauthoryear{{Schlafly} \& {Finkbeiner}}{{Schlafly} \&
  {Finkbeiner}}{2011}]{schlafly11}
{Schlafly} E.~F.,  {Finkbeiner} D.~P.,  2011, \mn@doi [\apj]
  {10.1088/0004-637X/737/2/103}, \href
  {https://ui.adsabs.harvard.edu/abs/2011ApJ...737..103S} {737, 103}

\bibitem[\protect\citeauthoryear{{Schlegel}, {Finkbeiner}  \&
  {Davis}}{{Schlegel} et~al.}{1998}]{schlegeldust}
{Schlegel} D.~J.,  {Finkbeiner} D.~P.,   {Davis} M.,  1998, \mn@doi [\apj]
  {10.1086/305772}, \href
  {https://ui.adsabs.harvard.edu/abs/1998ApJ...500..525S} {500, 525}

\bibitem[\protect\citeauthoryear{{Simon}}{{Simon}}{2019}]{simon19}
{Simon} J.~D.,  2019, \mn@doi [\araa] {10.1146/annurev-astro-091918-104453},
  \href {https://ui.adsabs.harvard.edu/abs/2019ARA&A..57..375S} {57, 375}

\bibitem[\protect\citeauthoryear{{Simon} \& {Geha}}{{Simon} \&
  {Geha}}{2007}]{simon07}
{Simon} J.~D.,  {Geha} M.,  2007, \mn@doi [\apj] {10.1086/521816}, \href
  {https://ui.adsabs.harvard.edu/abs/2007ApJ...670..313S} {670, 313}

\bibitem[\protect\citeauthoryear{{Stetson}}{{Stetson}}{1987}]{stetson87}
{Stetson} P.~B.,  1987, \mn@doi [\pasp] {10.1086/131977}, \href
  {https://ui.adsabs.harvard.edu/abs/1987PASP...99..191S} {99, 191}

\bibitem[\protect\citeauthoryear{{Stetson}}{{Stetson}}{1994}]{stetson94}
{Stetson} P.~B.,  1994, \mn@doi [\pasp] {10.1086/133378}, \href
  {https://ui.adsabs.harvard.edu/abs/1994PASP..106..250S} {106, 250}

\bibitem[\protect\citeauthoryear{{Tollerud}, {Bullock}, {Strigari}  \&
  {Willman}}{{Tollerud} et~al.}{2008}]{tollerud08}
{Tollerud} E.~J.,  {Bullock} J.~S.,  {Strigari} L.~E.,   {Willman} B.,  2008,
  \mn@doi [\apj] {10.1086/592102}, \href
  {https://ui.adsabs.harvard.edu/abs/2008ApJ...688..277T} {688, 277}

\bibitem[\protect\citeauthoryear{{Tollerud}, {Geha}, {Grcevich}, {Putman},
  {Weisz}  \& {Dolphin}}{{Tollerud} et~al.}{2016}]{tollerud16a}
{Tollerud} E.~J.,  {Geha} M.~C.,  {Grcevich} J.,  {Putman} M.~E.,  {Weisz}
  D.~R.,   {Dolphin} A.~E.,  2016, \mn@doi [\apj] {10.3847/0004-637X/827/2/89},
  \href {https://ui.adsabs.harvard.edu/abs/2016ApJ...827...89T} {827, 89}

\bibitem[\protect\citeauthoryear{{Torrealba}, {Koposov}, {Belokurov}  \&
  {Irwin}}{{Torrealba} et~al.}{2016}]{torrealba16a}
{Torrealba} G.,  {Koposov} S.~E.,  {Belokurov} V.,   {Irwin} M.,  2016, \mn@doi
  [\mnras] {10.1093/mnras/stw733}, \href
  {https://ui.adsabs.harvard.edu/abs/2016MNRAS.459.2370T} {459, 2370}

\bibitem[\protect\citeauthoryear{{Torrealba} et~al.,}{{Torrealba}
  et~al.}{2018}]{torrealba18}
{Torrealba} G.,  et~al., 2018, \mn@doi [\mnras] {10.1093/mnras/sty170}, \href
  {https://ui.adsabs.harvard.edu/abs/2018MNRAS.475.5085T} {475, 5085}

\bibitem[\protect\citeauthoryear{{Torrealba} et~al.,}{{Torrealba}
  et~al.}{2019}]{torrealba19}
{Torrealba} G.,  et~al., 2019, \mn@doi [\mnras] {10.1093/mnras/stz1624}, \href
  {https://ui.adsabs.harvard.edu/abs/2019MNRAS.488.2743T} {488, 2743}

\bibitem[\protect\citeauthoryear{{Walsh}, {Willman}  \& {Jerjen}}{{Walsh}
  et~al.}{2009}]{walsh09}
{Walsh} S.~M.,  {Willman} B.,   {Jerjen} H.,  2009, \mn@doi [\aj]
  {10.1088/0004-6256/137/1/450}, \href
  {https://ui.adsabs.harvard.edu/abs/2009AJ....137..450W} {137, 450}

\bibitem[\protect\citeauthoryear{{Weisz}, {Dolphin}, {Skillman}, {Holtzman},
  {Gilbert}, {Dalcanton}  \& {Williams}}{{Weisz} et~al.}{2014}]{weisz14}
{Weisz} D.~R.,  {Dolphin} A.~E.,  {Skillman} E.~D.,  {Holtzman} J.,  {Gilbert}
  K.~M.,  {Dalcanton} J.~J.,   {Williams} B.~F.,  2014, \mn@doi [\apj]
  {10.1088/0004-637X/789/2/147}, \href
  {https://ui.adsabs.harvard.edu/abs/2014ApJ...789..147W} {789, 147}

\bibitem[\protect\citeauthoryear{{Willman} et~al.,}{{Willman}
  et~al.}{2005a}]{willman05a}
{Willman} B.,  et~al., 2005a, \mn@doi [\aj] {10.1086/430214}, \href
  {https://ui.adsabs.harvard.edu/abs/2005AJ....129.2692W} {129, 2692}

\bibitem[\protect\citeauthoryear{{Willman} et~al.,}{{Willman}
  et~al.}{2005b}]{willman05b}
{Willman} B.,  et~al., 2005b, \mn@doi [\apjl] {10.1086/431760}, \href
  {https://ui.adsabs.harvard.edu/abs/2005ApJ...626L..85W} {626, L85}

\makeatother
\end{thebibliography}








\bsp	
\label{lastpage}
\end{document}